# THE SPALLATION NEUTRON SOURCE (SNS) LINAC RF SYSTEM*

Michael Lynch, William Reass, Daniel Rees, Amy Regan, Paul Tallerico,
Los Alamos National Laboratorys, Los Alamos, NM, 87544, USA

*Abstract*

The SNS is a spallation neutron research facility being built at Oak Ridge National Laboratory in Tennessee [1]. The Linac portion of the SNS (with the exception of the superconducting cavities) is the responsibility of the Los Alamos National Laboratory (LANL), and this responsibility includes the RF system for the entire linac. The linac accelerates an average beam current of 2 mA to an energy of 968 MeV. The linac is pulsed at 60 Hz with an H$^-$ beam pulse of 1 ms. The first 185 Mev of the linac uses normal conducting cavities, and the remaining length of the linac uses superconducting cavities [2]. The linac operates at 402.5 MHz up to 87 MeV and then changes to 805 MHz for the remainder. This paper gives an overview of the Linac RF system. The overview includes a description and configuration of the high power RF components, the HV converter/modulator, and the RF controls. Issues and tradeoffs in the RF system will be discussed, especially with regards to the use of pulsed superconducting cavities.

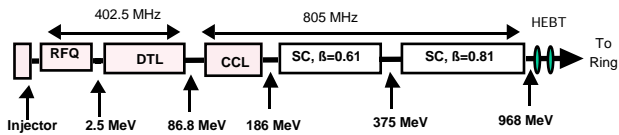

Figure 1: SNS Linac configuration.

## 1 OVERVIEW OF SNS LINAC

The SNS Linac is a nominal 968 MeV, 2 mA average, H$^-$ accelerator (Figure 1). The system provides 52 mA peak at 1 ms pulse width and a 60 Hz repetition rate. The beam is chopped with a 68% chopping factor, for an average beam during the pulse of 36 mA. The accelerator begins with an injector and RFQ (from Berkeley), then a Drift Tube Linac (DTL) to 86.8 MeV, followed by a Coupled-Cavity Linac (CCL) to 186 MeV. Los Alamos is responsible for the DTL and CCL structures and all of the linac RF systems. The remaining part of the accelerator uses superconducting cavities (from Jefferson Laboratory), using ß=0.61 cavities to 375 MeV, then ß=0.81 cavities to 968 MeV. The RFQ and DTL operate at 402.5 MHz, and the CCL and superconducting cavities operate at 805 MHz.

The different structures and frequencies require that 3 different types of klystrons be used. The RFQ and DTL will use 2.5 MW peak klystrons at 402.5 MHz. The CCL
_________________________________________________________
*Work supported by the US Department of Energy

will use 5 MW peak klystrons at 805 MHz, and the superconducting cavities will use 550 kW klystrons at 805 MHz. The types and quantities and applications are listed in Table 1.

Table 1: Types and quantities of klystrons and HV systems for the SNS Linac

| | |
|---|---|
| H$^-$ Energy | 968 MeV |
| Average beam during pulse | 36 mA |
| Pulse Width | 1 ms |
| Rep Rate | 60 Hz |
| **Klystrons** 402.5 MHz, 2.5 MW pk *(includes 1 for RFQ, 6 for DTL)* | 7 |
| 805 MHz, 5 MW pk *(includes 4 for CCL, 2 for HEBT)* | 6 |
| 805 MHz, 0.55 MW pk, SC | 92 |
| HV Converter/Modulators (16 Total) | 1 for each 5 MW klystron or pair of 2.5 MW klystrons *(except 1 for RFQ and first 2 DTL's & 1 for 2 HEBT cavities)* 1 for 11 or 12 0.55 MW klystrons |

Figure 2 shows a block diagram of the RF systems for the superconducting portion of the accelerator. Depending on the power required in the particular portion of the linac, each grouping consists of either 11 or 12 klystrons per HV system.

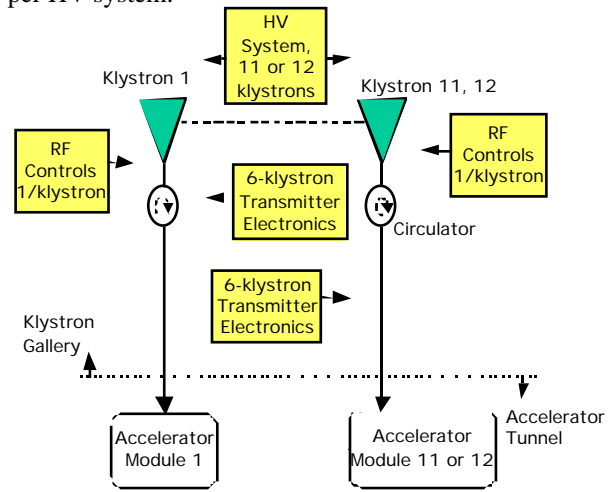

Figure 2: RF System block diagram for superconducting cavities. For room temperature cavities, each klystron has its own RF controls and Transmitter electronics.

In order to standardize as much as possible, the HV systems are designed to provide power to klystrons for 'approximately' 5 MW of RF. Therefore, the 402.5 MHz klystrons are configured with 2 klystrons per power supply. The CCL uses one klystron per HV system, and the superconducting klystrons are combined in groups of 11 or 12 per HV system (depending on the amount of power needed for that portion of the linac). Transmitters provide support electronics and interlocks to 2 klystrons (402.5 MHz), 1 klystron (CCL), or 6 klystrons (superconducting systems). Each klystron along the entire linac has its own feedback/feedforward RF control system. The power required along the superconducting portion of the linac is shown in Figure 3. Since the SNS is a proton machine, the power required in each cavity varies, depending on the cavity ß and the velocity of the particles. Since one HV supply provides power to 11 or 12 klystrons, each klystron in that group has the same saturated power level, set by the highest power requirement in the group.

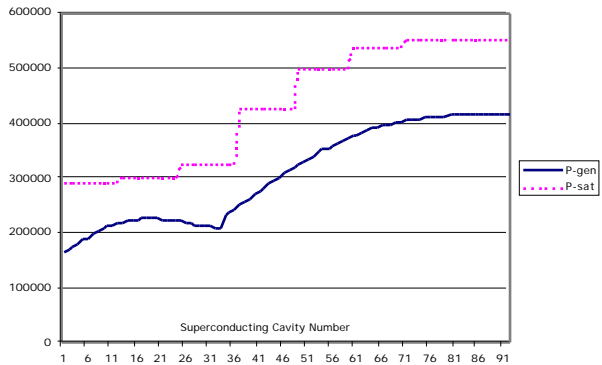

Figure 3: The required power level and the saturated power setting for each klystron in the superconducting portion of the SNS Linac.

## 2 RF SYSTEM COMPONENTS

### 2.1 Transmitter

The transmitter includes the klystron support tank and all of the support and interlock electronics for the high power RF systems. The transmitter includes the HV metering, the klystron filament power supply, the solid state driver amplifier, the klystron vac-ion pump power supply, the solenoid supplies, the interlocks for water flow, water temperature, and air cooling, the accelerator window cooling diagnostics, the AC distribution, and the user interface, consisting of the programmable controller and user display.

### 2.2 High Voltage (HV) System

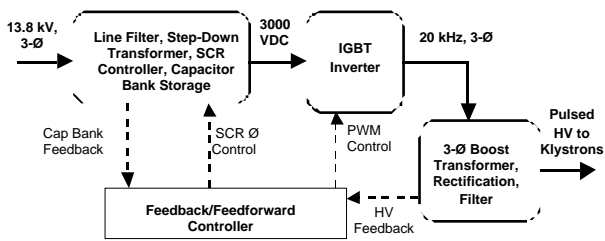

Figure 4: HV Converter/Modulator

The HV System provides everything from the input 13.2 kVAC line to the klystron, including the electronics to provide pulsed HV to the klystron cathode. Los Alamos is developing a unique design that provides conversion of HV-AC to HV-DC as well as the high voltage pulse modulation needed for SNS [3]. The design is very conservative with space and appears to be a very cost-effective approach as well. A block diagram of the converter-modulator is shown in Figure 4. The design uses standard 60 Hz technology to convert the incoming 13.2 kVAC to 3 kVDC. It then uses IGBT's, operating at 20 kHz switching speeds into a 3-phase inverter step-up circuit and a high voltage rectifier/filter, to convert the low voltage DC into the required pulsed, high voltage DC for the klystrons. The system is very versatile in that the only changes needed between the 5 MW klystron application (135 kV, 70 A) and the 550 kW klystron application (75 kV, 130 A) are the turns-ratio and peaking-capacitor value in the 3-phase high voltage step-up transformer. The system employs pulse-width modulation (PWM) of the IGBT switching to maintain regulation. The control of this PWM as well as the SCR phase control is achieved by a combination of both feedback and feedforward.

### 2.3 RF Controls

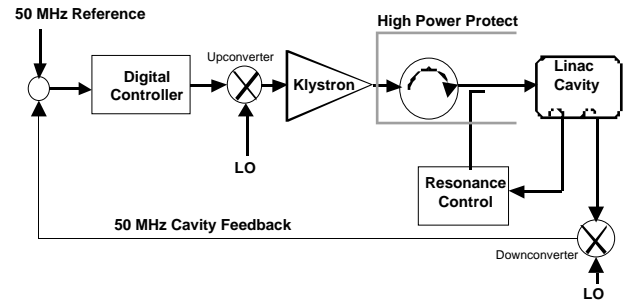

Figure 5: RF Controls for SNS

The RF control system has to control RF systems consisting of both room temperature and superconducting cavities, frequencies of both 402.5 and 805 MHz, and klystrons of 2.5 MW, 5 MW, and 550 kW [4,5]. The control system is based on the controls originally developed for the Ground Test Accelerator at LANL, and then recently adapted for the Accelerator for Production of Tritium. The system is based on VXI hardware. A block diagram of the control system is shown in Figure 5. The original GTA system did all of the actual control electronics in analog circuitry. For APT , the system used a combination of digital (for low frequency control) and analog (for higher speed control). Recent advances in the speed of digital electronics has led us to pursue an all-digital control system for use on SNS. In addition it will make extensive use digital signal processors (DSP's) which will allow changes in the control algorithms by reprogramming the DSP's rather than changing the hardware.

We have also developed an extensive model of the control system, including such things as klystron saturation, loop delay, microphonics, Lorentz-force detuning, and beam and power supply noise [6]. The model has been developed in MATLAB/SIMULINK, and allows us to optimize the design approach before producing any hardware. An example of one modeling result is shown in Figure 6.

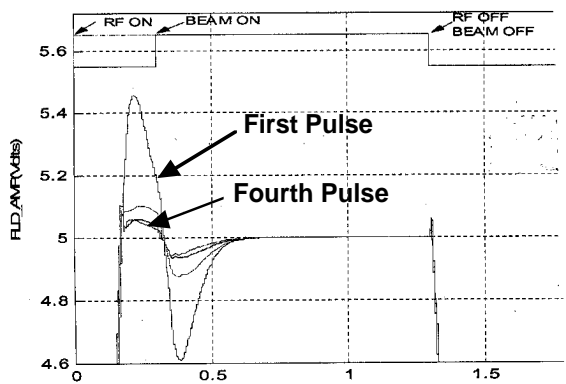

Figure 6: Modeling results for SNS superconducting cavity showing the effect of iterative learning control. Each successive pulse shows a reduced error due to improved feedforward correction.

## 3 PULSED OPERATION OF A SUPERCONDUCTING PROTON ACCELERATOR

Until recently, superconducting cavities have been used only in CW applications. Pulsed operation of a superconducting accelerator has been addressed recently in work at DESY on the Tesla facility, in Japan for the Joint Project, and at various places for the European Spallation Source (ESS). The SNS is adding to that work and will make advances especially with respect to the use of pulsed superconducting cavities for proton acceleration. The original superconducting design for SNS had external Q's of approximately 5e5 and $E_{acc}$ of about 11.9 MV/m. The resultant effect of Lorentz detuning and microphonics was relatively small. As the design of SNS has progressed, however, the external Q's have increased (to about 7e5) and $E_{acc}$ has increased (to 14.2 MV/m). This means that the effects of Lorentz detuning and microphonics, while not at the Tesla level, have become a more serious problem. Current external bandwidths are about 550 Hz, and the expected microphonics and Lorentz detuning total is 400 to 500 Hz.

Because it is a proton accelerator, SNS has the added problem of nonzero synchronous phase (to maintain longitudinal bunching). In addition, the synchronous phase of the machine varies from one cavity to the next (over a range of $-26.5°$ to $-15°$). To minimize the effect of reflected power, one typically detunes the cavity so that the reactive effect of the nonzero synchronous phase is cancelled by the cavity detuning. With Lorentz forces and microphonics, it is impossible to maintain exact cavity detuning. The system must therefore be designed to handle the reflected power, both in terms of the RF components and the power capability in the RF source.

## 4 TECHNOLOGY DEVELOPMENT AND ALTERNATIVE CONFIGURATIONS

Other superconducting applications have spent several years in technology development activities. The SNS is scheduled to begin commissioning in 2005, with first operation in 2006. There is little time for development. All 105 RF systems must be designed, built, integrated, installed, and commissioned in only 5 Years. This puts great emphasis on using what is available, on designing a robust system that can accommodate changes, and on using dependable models for the development.

Many different configurations of the linac and RF systems were considered in terms of the cost and performance. The current baseline, as presented in this paper, represents the necessary technology advancements to meet the performance requirements while staying within the very tight budget and schedule limitations.

In addition, as in most modern accelerator applications, there is always the need to consider other applications for the system in order to make best use of the large capital investment. In addition to the basic 60 Hz application, we are considering the repercussions of interleaving a 10 Hz pulse pattern for a second neutron target. This has implications for the power generation and handling, for the feedforward control systems, and for the definition of component requirements as we begin making the procurements. The primary consideration at this point is to ensure that we do not preclude the option of the interleaved pulses and the second target.